# Transfer of Graphene with Protective Oxide Layers

H. Grebel[1], L. Stan[2], A. V. Sumant[2], Y. Liu[2], D. Gosztola[2], L. Ocola[2] and B. Fisher[2]

[1]*Electronic Imaging Center and ECE Dept., New Jersey Institute of technology (NJIT), Newark, NJ 07102, USA. grebel@njit.edu*

[2]*Center for Nanoscale Materials, Nanoscience and Technology Division, Argonne National Laboratory Argonne, IL 60439*

**Abstract:** Transfer of graphene, grown by Chemical Vapor Deposition (CVD), to a substrate of choice, typically involves deposition of a polymeric layer (typically, poly(methyl methacrylate, PMMA or polydimethylsiloxane, PDMS). These polymers are quite hard to remove without leaving some residues behind. Here we study a transfer of graphene with a protective thin oxide layer. The thin oxide layer is grown by Atomic Deposition Layer (ALD) on the graphene right after the growth stage on Cu foils. One can further aid the oxide-graphene transfer by depositing a very thin polymer layer on top of the composite (much thinner than the usual thickness) following by a more aggressive polymeric removal methods, thus leaving the graphene intact. We report on the nucleation growth process of alumina and hafnia films on the graphene, their resulting strain and on their optical transmission. We suggest that hafnia is a better oxide to coat the graphene than alumina in terms of uniformity and defects.

## I. Introduction:

Typical transfer of graphene - grown by CVD on Cu or Ni substrates - onto a substrate of choice, involves some intermediate stages [1-9]. The graphene-deposited metal substrate is coated with a polymer, e.g., poly(methyl methacryslate) (PMMA) [1], or polydimethylsiloxane (PDMS) [2] in order to protect it through the following transfer stages. The metal films are etched away and the graphene/polymer film is transferred and let rested onto the desired substrate. The polymer is later removed by dissolving it out, or peeling it off. Any aggressive polymer removal method, such as reactive ion etching, is prohibited for fear of damaging the graphene layer. The polymer is problematic because remnants of the polymer may be left behind and impede the following processing stages. At the same time, many applications, e.g., graphene based channel transistors, require a top gate oxide layer. One would wonder, then, whether the graphene may be protected by a top oxide layer throughout the transfer stages. Here we study a transfer of a CVD grown graphene with a layer of protective oxide: alumina ($Al_2O_3$) and hafnia ($HfO_2$) deposited by Atomic Layer Deposition (ALD).

The growth of the oxide on the graphene poses some problems. Pin-holes, poor adhesion and non-uniformity are attributed to the chemically inert graphene surface because the surface lacks dangling bonds to assist the adhesion process. A two-step oxide growth method was proposed [10-12] to overcome this impediment. This method was used to transfer relatively thick films (on the order of 40 nm, [13]) but direct view of the initial growth of thin layers with SEM and TEM and the optical transmission through oxide-on-graphene has been lacking. Here we study thin oxide films on graphene before and after the transfer stage to glass substrates.

## II. Experiment and initial Characterization:

The 25 micron thick Cu foil substrate, used in the graphene growth process has a pattern, remnant of its manufacturing (Fig. 1). Graphene was grown by using commercial Thermal CVD system (Atomate Inc.). The process involved annealing a Cu foil in hydrogen atmosphere prior to growth followed by graphene growth at high temperature (1000 $^o$C) using methane flow (50 sccm) as a carbon source at intermediate pressure (1 torr) for few minutes and cooling from growth temperature to room temperature in hydrogen atmosphere (300 Torr). This resulted in mostly bi-layer graphene on Cu foil [14] albeit Raman spectra reveals areas with single layer, as well. The template in the Cu foil during extrusion could result in further straining of the graphene layer. In principle, one may use flat Ni or Cu films on Si/$SiO_2$, instead [2].

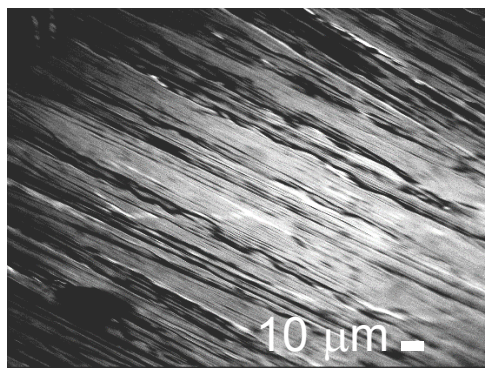

Fig. 1. CVD grown graphene on a 25 microns Cu foil. One may observe the Cu template due to the metal extrusion.

Aluminum oxide (alumina, $Al_2O_3$) and hafnium oxide (hafnia, $HfO_2$) were deposited by Atomic Layer Deposition The $Al_2O_3$ and $HfO_2$ films were grown on graphene by ALD in an Arradiance Gemstar system using a two-step growth process [10-12]. For the $Al_2O_3$ films, trimethylaluminum (TMA, STREM Chemicals) was used as a precursor and water as an oxidant and for the $HfO_2$ films, tetrakis(ethylmethylamino)hafnium (TEMAH, STREM Chemicals) was used as a precursor and $H_2O$ as an oxidant. The TEMAH precursor was heated to 95°C with a heating jacket while the water was kept at room temperature. High purity nitrogen gas (99.999%) was used as a carrying and purging gas with a flow rate of 40 sccm and 100 sccm, respectively. During the high exposure-mode ALD process, the evacuation valve was closed before introducing the precursor in the chamber and opened after a delay time. In the first step, a seed layer (3 cycles for both, $Al_2O_3$ and $HfO_2$) was grown at lower temperature (90 °C). During the seed layer growth, 3 pulses of water, 0.022s each were followed by 1 pulse of the metal precursor (0.022 s TMA and 1 s TEMAH, respectively). The usage of more water than for a conventional ALD process during this low temperature step, was intended to facilitate the adherence of water to the graphene thru physical absorption [11]. The second step was the bulk layer growth (100 cycles) at higher temperature (200 °C). An $Al_2O_3$ cycle consisted of 0.022s TMA pulse, 1 s delay, 28 s $N_2$ purge, 0.022 s $H_2O$ pulse, 1 s delay, and 28 s $N_2$ purge. An $HfO_2$ cycle consisted of 1 s TEMAH pulse, 2 s delay, 28 s $N_2$ purge, 0.022 s $H_2O$ pulse, 1 s delay, and 34 s $N_2$ purge. The resulting $Al_2O_3$ and $HfO_2$ films had a thickness of about 10 nm.

SEM pictures of 10-nm alumina and hafnia on bare Cu are shown in Fig. 2. The coverage of the oxide film is rather uniform on the Cu foil. This is probably because the copper is partially oxidized by the water, enabling a more homogeneous oxide growth. In contrast, the dangling bonds, provided by the hydroxide on graphene are more sporadic leading to oxide layer nucleation, islanding and wire-like growth.

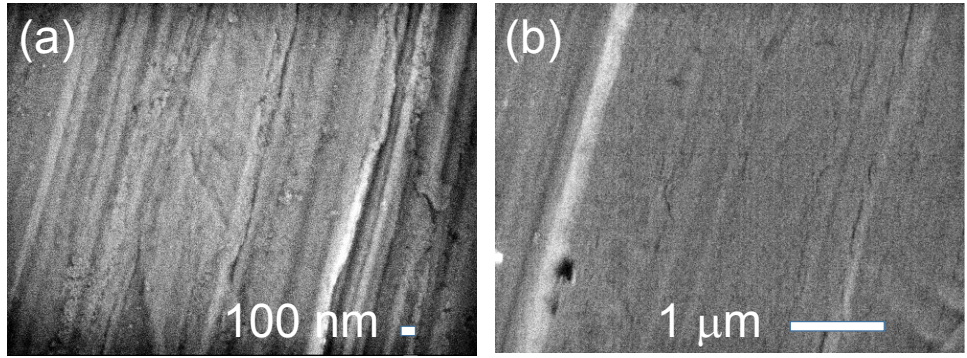

Fig. 2.  10-nm thick alumina (a) and 10-nm thick hafnia (b) on bare 25 micron thick Cu foil.

SEM pictures of 10-nm alumina and hafnia on graphene-coated Cu foils are shown in Fig. 3. The pictures reveal some pin-holes, which are more prevalent in alumina-graphene coatings than in hafnia-graphene films.

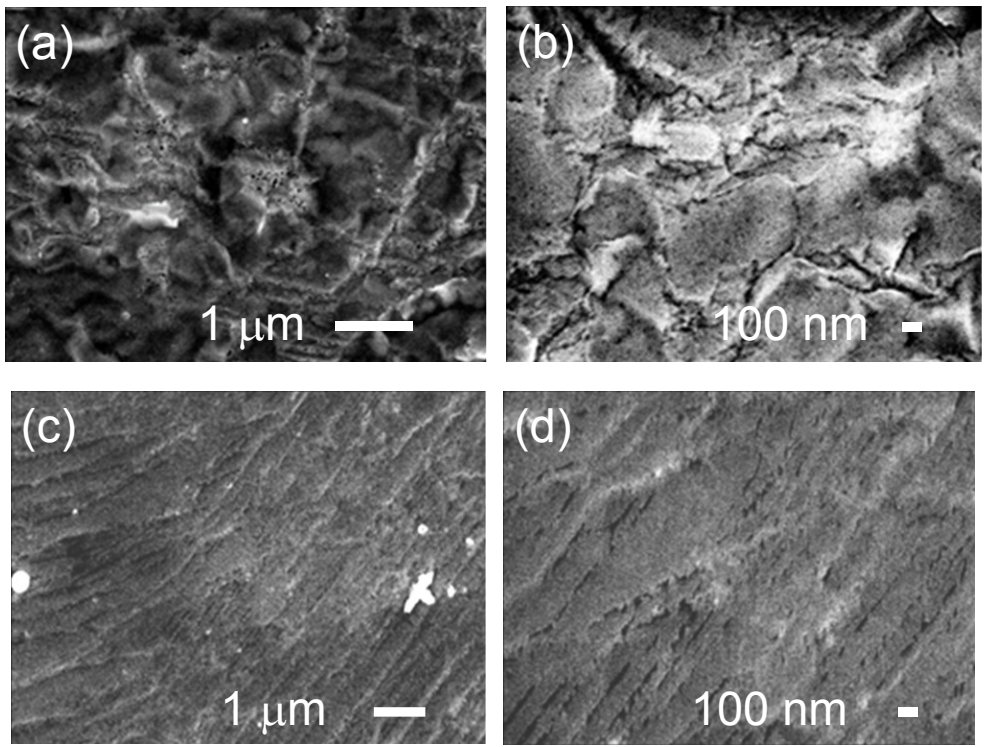

Fig. 3.  10-nm thick alumina (a,b) and 10-nm thick hafnia (c,d) on graphene coated 25 microns thick Cu foil.

Elemental study via EDS for aluminum-graphene films, as well as hafnia-graphene films, both deposited on Cu foil, are presented in Fig. 4.  Hafnia-graphene films on Cu exhibited a small amount of Al; such Al presence has been attributed to the alumina filters used for the water.

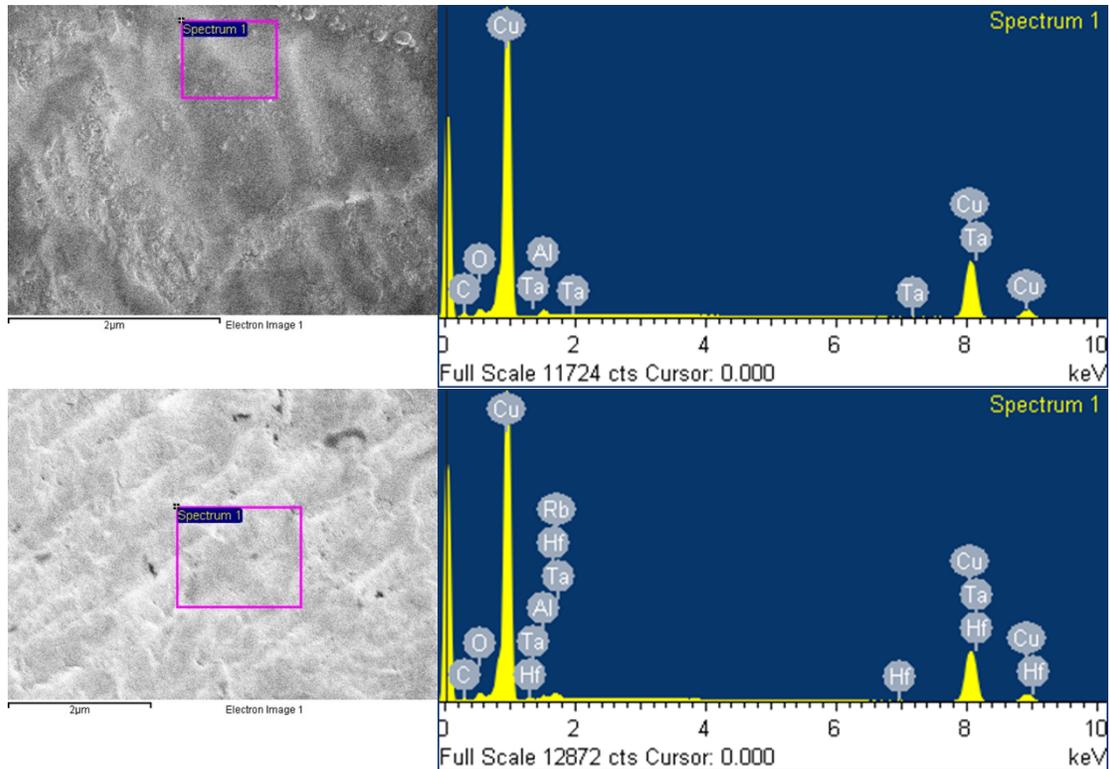

Fig. 4. Top: EDS of alumina-graphene film. Bottom: EDS of hafnia-graphene film.

TEM pictures exhibit a crystalline features superimposed on an overall amorphous circles (Fig. 5a,b). The 10-hafnia-graphene gave rise to a sharper diffraction pattern. One may note an unusual nucleation in the oxide-graphene films which could result in the bright diffraction spots. Such process may be understood as a two-component nucleation [15] (originally suggested by [16]) which starts at the OH⁻ sites on the graphene. In both Figs. 5, one may observe the formation of domains.

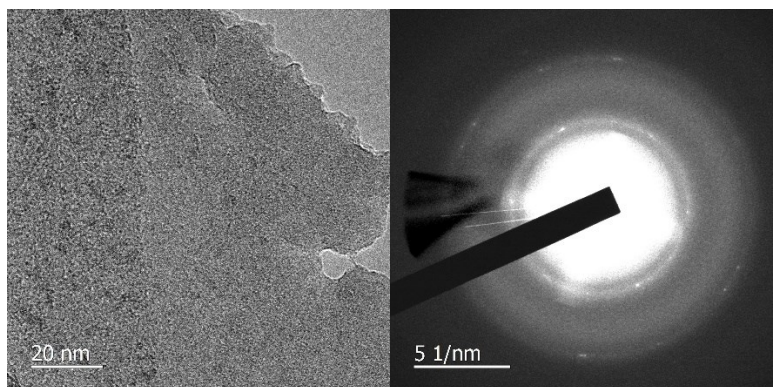

Fig. 5a. TEM image and diffraction pattern of 10-nm thick alumina-graphene film, suspended on a TEM grid. The bright spots in the diffraction pattern allude to nucleation of a two-component material.

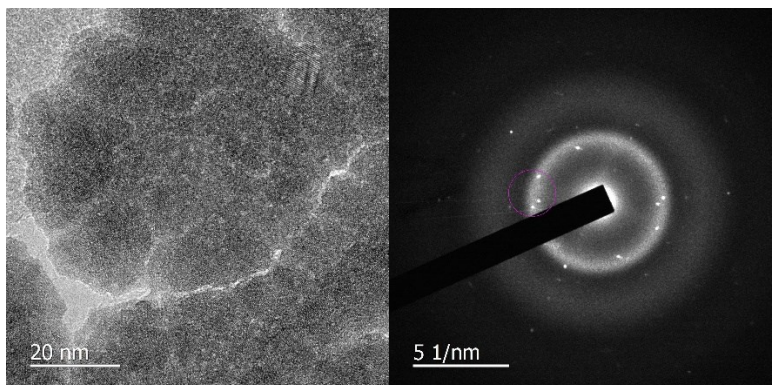

Fig. 5b. TEM image and diffraction pattern of 10-nm hafnia-graphene film, suspended on a TEM grid. The bright spots in the diffraction pattern allude to nucleation of a two-component material.

Raman spectra of 10-nm alumina-graphene and 10-nm hafnia-graphene both on 25 Cu foil by using the 442 nm HeCd laser line are shown in Fig. 6. The 2D line shift for the alumina-graphene as well as the shift in the G-line ought to be attributed to a 2-layer graphene; thus, both films seem to exhibit minimal stress on the copper foil. The two lines below 1000 1/cm belong to the Cu foil.

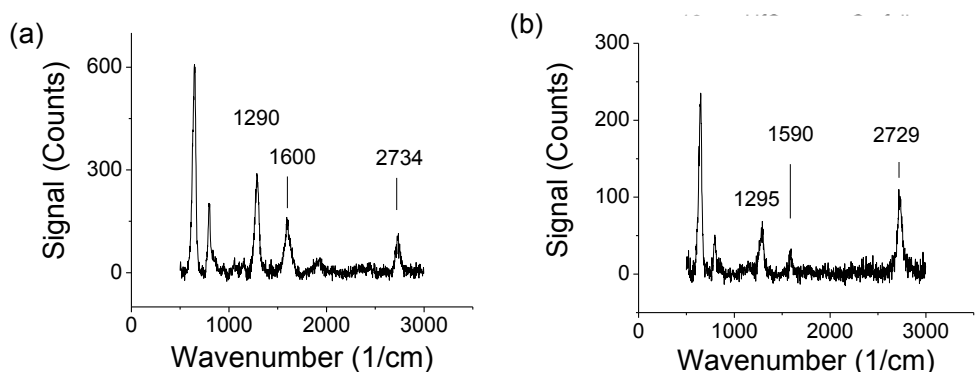

Fig. 6. (a) 10-nm alumina-graphene on a 25 micron Cu foil. (b) 10-nm hafnia-graphene on 25 micron Cu foil. The two lines below 1000 1/cm belong to the Cu foil itself. The 1290 1/cm is the defect D line when exposed to the 442 nm laser line. Local stresses and differences between 2- and 1-layer graphene may have contributed to the graphene's line shifts in towards the higher frequencies in (a) vs (b).

### III. Transferred Films:

In the following, we separate the discussion into oxide-graphene transfer without and with an additional step of a 50 nm PMMA layer. The PMMA thickness was estimated by depositing it on a 50 nm $SiO_2$ on a Si wafer; the additional PMMA layer contributed to a visible blue shade on top of the thermally oxidized wafer. In Fig. 7, we show an optical images of a transfer without a PMMA layer, which we refer to as a direct transfer. Unlike using only PMMA in the transfer of graphene, the original Cu template is visible after depositing it with an oxide.

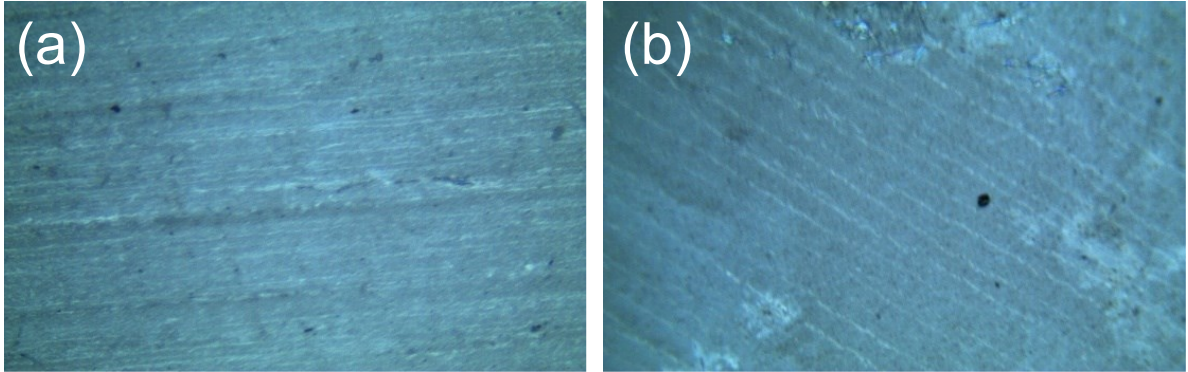

Fig. 7. Optical images of transferred graphene on glass. (a) 10-alumina-graphene on glass. (b) 10-nm hafnia-graphene on glass. The microscope objective was x20.

Related Raman spectra of the directly transferred alumina-graphene and hafnia-graphene onto glass are shown in Fig. 8. The films are the same as those studied in Figs. 6. The 442 nm laser line of a HeCd laser was used to eliminate as much as possible signal from the glass slides (since the shorter wavelength has a smaller penetration depth). As clearly visible from Fig. 8, the transferred alumina-graphene film gave rise to additional graphene lines, most notably at 1389 and 1478 1/cm. The first may be attributed to a modified graphitic defect line while the latter may be attributed to the effect of nitrogen. The modified defect line is missing from the data of alumina-graphene on Cu. This line ought to be correlated with the down-shift of the G-line which indicates strain. Clustering of graphene leads to a D line intensity increase [17]. The 1478 1/cm is typical of aromatic rings which could be formed around the $OH^-$ defect point in the graphene with the nitrogen in the alumina and adsorbed water molecules during the transfer stage. Strain at the defect point may distort the graphene π-bonds and lead to an enhanced reactivity (in this case water) even though the temperature involved is quite low (<200 °C) [18-19]. The defect line remained small in the hafnia-graphene layer and no additional line is observed at 1450-1500 1/cm. This is also true when one transfers graphene with PMMA without the alumina.

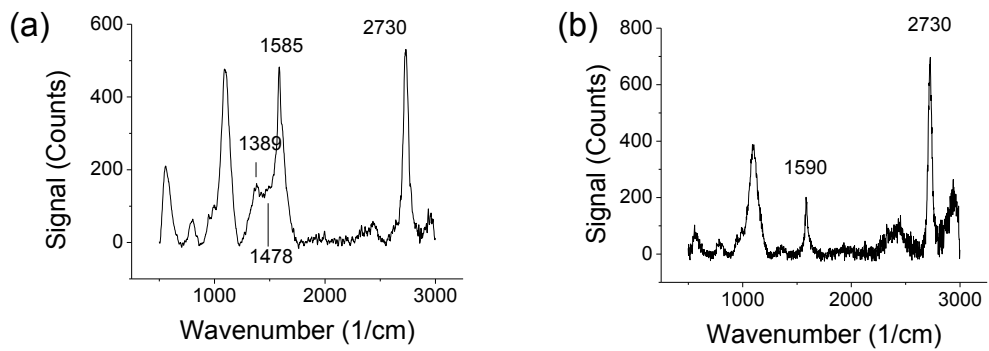

Fig. 8. Direct transfer of 10-nm oxide-graphene from Cu foil onto a glass substrate. (a) 10-nm thick alumina-graphene. (b) 10-nm thick hafnia-graphene. Spectra were taken with the 442 line of a HeCd laser of the Renishaw Raman system. The lines below 1200 1/cm are attributed to the glass slide.

One can also transfer the 10-nm oxide-graphene layer by depositing it with an additional, ca 50 nm thick PMMA film. The PMMA film is invisible on a glass slide but has a bluish taint when deposited on a 50-nm thick oxide on a Si wafer.  The addition of the PMMA film makes the transfer process much easier because the suspended oxide-graphene is less fragile.  In our case, we used acetone at 40 °C or Anisole at RT to remove the PMMA.  The PMMA film used was much thinner than a typical film used for graphene transfer – a 50-nm thick film compared with a typical thickness of 150-nm - and need not be heated thoroughly (less than 1 min. on a 180 °C hot plate or as low as 1 min on a 60 °C plate, compared with >3 min. on a 180°C plate).  Images of the transferred graphene using an additional 50-nm 495 A2 PMMA resist is shown in Fig. 9a.  The 495A2 PMMA resist (MicroChem) was later removed using Anisole for 30 min at RT.  The related Raman spectrum with a 442 nm laser is shown in Fig. 9b.  Two additional lines at 1336 and 1447 1/cm are observed.  The first is a modified graphitic defect line while the other may be attributed to aromatic ring as discussed earlier.  Note that the PMMA relieved some of the defect strain: the G-line has not been down shifted while the modification in the D-line is smaller than the one exhibited in Fig. 8a.

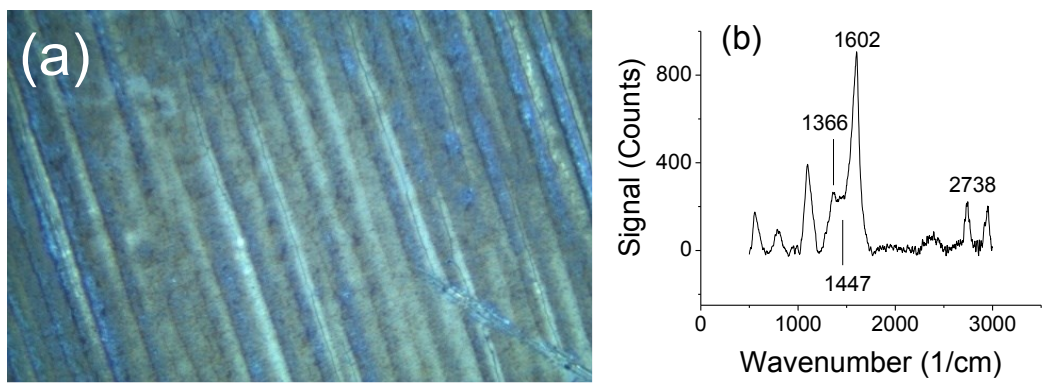

Fig. 9.  (a) An image of transferred 10-nm alumina-graphene on glass slide by using an additional step of coating the sample with 50-nm PMMA film and remove the polymer later.  (b) Raman spectra.  Spectra were taken with the 442 line of a HeCd laser of a Renishaw Raman system.

Electrical data are shown in Fig. 10.  The sample is shown in Fig. 10a. The transfer of the alumina-graphene film was aided by an additional layer of 50-nm thick PMMA.  The PMMA was later removed with acetone at 40 °C for 15 min.  We used the Si wafer as a back gate electrode. Keithley 236 system, capable of measuring current levels, as low as 100 fA was used in the measurements.  The pin-holes in the oxide layer enabled an ohmic contact of the silver paste to the graphene layer.  Ohmic contact is exhibited by the linear Ids-Vds curve and the Dirac point, observed at Vgs=0 V.  The back and forth scans led to very small error bars.  This demonstrates that the basic electrical properties of the graphene stayed intact.

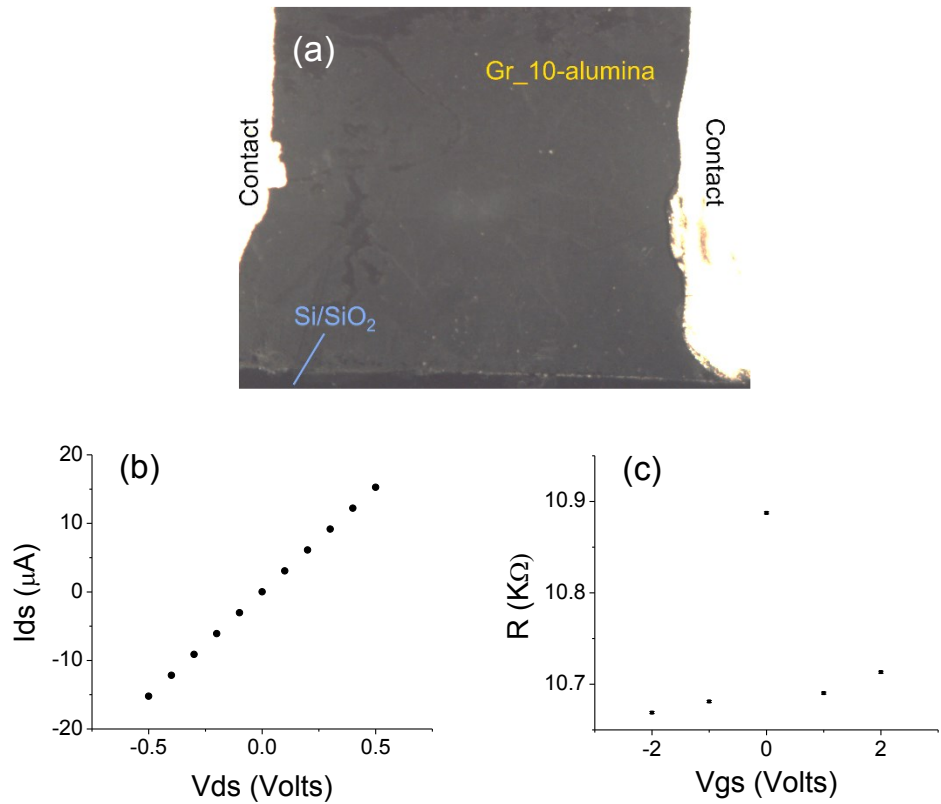

Fig. 10 (a) 10-nm alumina on graphene when transferred on a 50 nm thermal oxide on a Si wafer. The graphene area measured 7x6 mm2. Also shown are (b) Ids-Vds at Vgs=+2 V and (c) channel resistance as a function of Vgs. A 22 KOhm resistance of the probes has been subtracted from the data.

In Fig 11 we show SEM picture of 5-nm alumina on graphene while on the copper foil. Wires or bands of alumina on graphene showing various growth modes (Fig 11a vs Fig. 11b). It is tempting to relate the rhombic patterns in Fig. 11b to single crystalline structures yet, as we saw in Fig. 5a, the alumina-graphene is formed through nucleation whose domains could be initiated by the Cu template.

For the electrical measurements, the film measured 5x8 mm but the distance between the electrodes was 3 mm. Instead of putting silver paste electrodes on top as in Fig 10a, we let the graphene rest on silver epoxy electrodes (Ted-Pella) which were deposited directly on a 50-nm $SiO_2$ on a Si wafer (and hence were exposed to water and cleaning agents). Such process may have led to local charges as exhibited by the shift in the graphene's Dirac point.

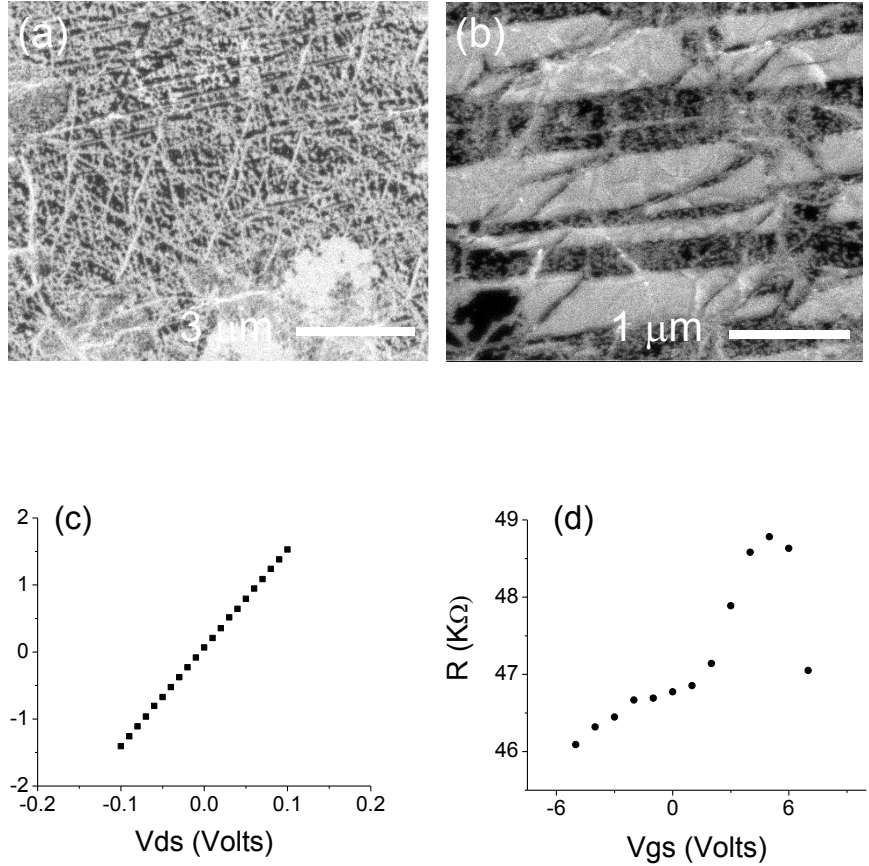

Fig. 11. (a,b) Various growth mode can be identified for 5-nm alumina on graphene when still on a 25 micron thick Cu foil. Lower panels: Ids-Vds at Vgs=-5 V for a transferred graphene on a 50-nm SiO$_2$ on Si and Channel Resistivity, R as a function of Vgs exhibited a Dirac point at Vgs=5 Volts.

Comparative Raman spectroscopy on a transferred bilayer graphene is shown in Fig. 12. Shown are two samples: one is deposited with 10-nm and the other with 5-nm of alumina. The transfer process involved the use of 50-nm coating of PMMA (see above). The Raman data, taken with a 532 nm laser indicate that the thicker oxide did not lend itself to a larger stress as judged by the 2D line. However, the increase in the D line and the small down-shift of the G-line point to strained defects.

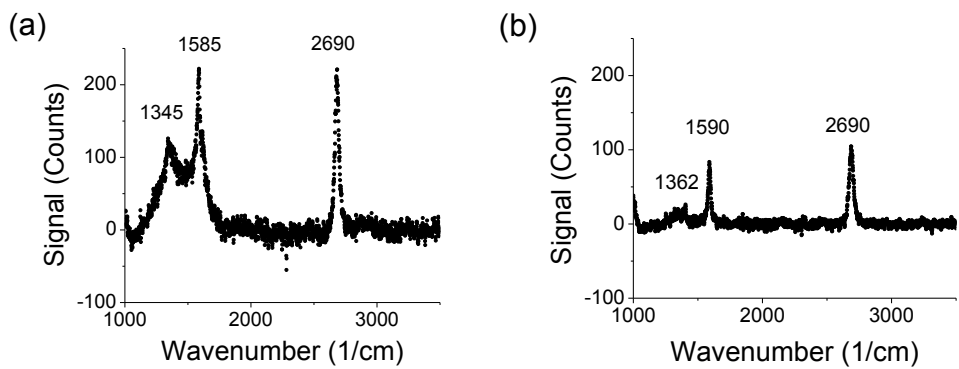

Fig. 12. Comparative Raman spectra of bilayer graphene deposited with alumina when transferred onto thermally prepared 50-nm $SiO_2$/Si. (a) 10-nm alumina on bi-layer graphene. (b) 5-nm alumina on bi-layer graphene. Raman data were obtained with a laser at 532 nm.

Optical transmission for transferred oxide-graphene films are presented in Fig. 13. All films were deposited on glass slides. The data were normalized by the transmission of only graphene on the glass. The samples were tilted a bit to minimize multi-reflection paths within the glass due to the relatively large refractive index of the oxide compared to the glass itself. The 10-nm alumina-graphene on a glass slide appeared darker to the naked eye probably because of scattering from the less homogeneous alumina film.

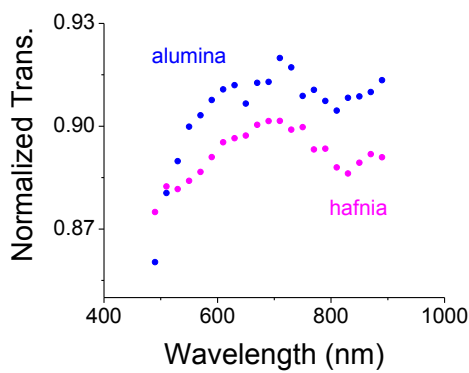

Fig. 13. Transmission of 10-nm oxide when depositing the graphene layer (on a glass slide). The data were normalized by the transmission of only graphene film on a glass slide.

## IV. Conclusions:

A study of transferred graphene with less than 10-nm of protective oxides revealed that the oxide layer did not contribute to a major film stress and preserved the electrical properties of the graphene. The oxide layer exhibited pin-holes. Raman data pointed to strained defect formation in alumina-graphene films but none in hafnia-graphene. Coating the oxide-graphene layer with an additional 50 nm of PMMA film made the transfer of graphene easy and its removal very efficient.

**Acknowledgment:** This work was performed, in part, at the Center for Nanoscale Materials, a U.S. Department of Energy Office of Science User Facility, and supported by the U.S. Department of Energy, Office of Science, under Contract No. DE-AC02-06CH11357. We thank X. Miao of NJIT for the EDS pictures. We also thank X. Miao for helping with a Raman scan.